\documentclass[english]{article}
\usepackage[T1]{fontenc}
\usepackage[koi8-r]{inputenc}
\usepackage{amstext}
\usepackage{esint}

\makeatletter
\@ifundefined{date}{}{\date{}}

\newtheorem{theorem}{Theorem}

\newtheorem{lemma}[theorem]{Lemma}

\usepackage{babel}

\makeatother

\usepackage{babel}
\begin{document}

\title{Fine structure of one-dimensional discrete point system}

\author{V. A. Malyshev}
\maketitle
\begin{abstract}
We consider the system of $N$ points on the segment of the real line
with the nearest-neighbor Coulomb repulsive interaction and external
force $F$. For the fixed points of such systems (fixed configurations)
we study the asymptotics (in $N$ and $l$) of finite differences
of order $l$. Classical theory of finite differences is extensively
used.
\end{abstract}
Assume that the system 
\[
0\leq x_{1}<x_{2}<...<x_{N}<1
\]
of $N$ different points on the segment $[0,1]\in R$ is given. If
this system is a random system, there exist a lot of ways to characterize
its structure, for example, as a random process of increments $\Delta_{i}=x_{i+1}-x_{i}$.
If there is no any randomness, then there is no conventional way to
characterize its organization. One of the possible ways is to consider
the system of finite differences of several orders $l=1,2,...$ 
\[
\nabla x_{i}=\Delta_{i}=x_{i+1}-x_{i},\nabla{}^{2}x_{i}=x_{i+2}-2x_{i+1}+x_{i},...,\nabla{}^{l}x_{i},...
\]
as the natural local characteristics of this system of points (or numbers),
even if we would not know the metrics of the space where they are
embedded and even would not know the space itself. This corresponds
to the situation in the analysis where the existence of derivatives
of sufficiently large order indicates the <<quality>> of the function.
In the discrete case there is no existence problem (for example, on
a circle the difference of arbitrary order are defined), but a possible
substitute for the existence can be the decay rate of finite differences
in $l$ (for sufficiently large $N$). To go further, everything depends
on the way how this systems of points is defined. Most popular way
is the discretization of smooth functions. Already for a long time
discrete differences have been used in numerical mathematics in connection
with approximation, interpolation, solution of differential equations,
etc. However, in these disciplines one does not need differences of
high orders. For arbitrary orders there exists classical science -
the theory of finite differences, see \cite{Jordan,Milne-Thomson,Gelfond}),
worked out already at Newton's time (the Newton series, divided differences,
etc,), In section 2 we give necessary definitions and results from
this science.

We are interested in the case when there is no natural smooth function
such that the point system is its discretization. Such systems appear,
for example, as the fixed points of natural dynamical systems in physics.

We suggest a new insight on the system of differences as the indicators
of the deviation scale from the ideal system. We call the system \textbf{ideal
system} if all distances between neighbors are equal. Then all differences
of the order greater than $1$ are zero. In physics this corresponds
to the ideal crystal, and many papers were devoted to the proof that
the system becomes ideal in the (thermodynamic) limit $N\to\infty$,
see \cite{Ventevogel_1,Duneau_01,Radin_01,Radin_02,Radin_04}. We
are interested in the cases when the system is (in some sense) close
to ideal. Here the thermodynamic limit is counter-productive - our
problem is finer.

The differences of given order $l$ depend on $N$ and on $i$. We
are interested in the asymptotics or, at least, in the bounds from
above for given $l$, uniform in $i$. Let us say that the differences
of order $l$ are defined on the scale $\kappa(N,l)$ (not seen on
the scales higher than $\kappa(N,l)$) , if the number $\kappa(N,l)$
is the asymptotics for the differences of order $l$ uniform in $i$
(bound from above, for given $l$, uniform in $i$). We call \textbf{fine
structure} of the point system the set of numbers $\kappa(N,l)$.

For example, in physics the scale of order $1$ corresponds to macro-scale,
and the scale $N^{-1}$ corresponds to the micro-scale.

\paragraph{Formulation of the problem}

Often, the system of point is not given explicitly, but as a configuration,
yielding minimum to some potential, or as a fixed point of some dynamics.
We study here concrete system of points on $[0,1]$, defined by the
system of equations 
\begin{equation}
f(x_{i}-x_{i-1})-f(x_{i+1}-x_{i})+F(x_{i})=0,i=2,...,N-1\label{MainEquations}
\end{equation}
and for any end point $x_{1}$ and $x_{N}$ there is an alternative
- either $x_{1}=0$ ($x_{N}=1$) and moreover correspondingly 
\[
-f(x_{2}-x_{1})+F(x_{1})\leq0\,\,(f(x_{N}-x_{N-1})+F(x_{N})\geq0)
\]
or, correspondingly, 
\[
-f(x_{2}-x_{1})+F(x_{1})=0\,\,(f(x_{N}-x_{N-1})+F(x_{N})=0)
\]
This system is interpreted as a fixed configuration, when $x_{i}$
are subjected to the interaction forces $f(x_{i}-x_{i-1})$ and $-f(x_{i+1}-x_{i})$
with the left and right neighbor correspondingly (we consider the
Coulomb repulsive interaction $f(r)=r^{-2}$). Moreover, there is
external force $F(x)$. It is clear that if $F(x)\equiv0$, then $x_{1}\text{=}0,x_{N}=1$,
and all $x_{i+1}-x_{i}$ are equal - the ideal case.

In this paper, which is a natural continuation of the papers \cite{Mal-1,Mal-2},
we consider three examples of the external force $F$ - constant,
linear and power functions. Only in the first case we get exact asymptotics,
which is based on the techniques, developed in combinatorics for Stirling
numbers of the second kind. In two other cases we get only bounds
from above, which seem to be close to exact, as our estimates seem
to be <<on the edge>> of exact estimates. Proof, in the linear case,
are based again on the combinatorics and inductive procedure in $l$.
In the power case we use a different inductive construction, however
each step of this construction contains one more inductive procedure,
similar to the one used in linear case.

It was proved in \cite{Mal-1,Mal-2}, that for any monotone external
force $F(x)$ the fixed configuration exists, is unique and such that
$x_{1}=0,x_{N}=L$, and for $i=2,...,N-1$ the equations (\ref{MainEquations}).
Moreover, it was proved there that the first differences have asymptotics
(uniform in $i$) 
\[
\Delta_{i}\sim\frac{L}{N-1}
\]
The central fact - this asymptotics does not depend on $F$. In \cite{Mal-1}
the second term of the asymptotic expansion was obtained 
\[
(x_{i+1}-x_{i})-\frac{L}{N-1}\sim\frac{F}{2}N^{-3}(\frac{N}{2}-i),i=2,...,N-1
\]
for constant external force $F$. Formal calculation shows that the
main term of the asymptotics for the second difference should be of
the order $N^{-3}$, and does not depend on $i$, and moreover using
the results of \cite{Mal-1}, one can show that the higher terms of
the asymptotic expansion do not change this conclusion. One can say
that only second difference shows up the macro force $F$. Here we
study the asymptotics of all higher differences. Denote 
\[
\Delta_{1}=x_{2}-x_{1}=\Delta,\Delta_{i}=x_{i+1}-x_{i}=\Delta(1+\delta_{i}),i=1,...,N-1
\]
Thus $\delta_{1}=0$ by definition, and in \cite{Mal-1,Mal-2} it
was proved that $\Delta\sim\frac{1}{N}$. All proofs below are based
on the study of the following system of equations for the unknowns
$\delta_{i},i=2,...,N-1$

\[
f(\Delta(1+\delta_{i}))-f(\Delta)=\sum_{i=2}^{i}F(x_{i}),x_{i}=(i-1)\Delta+\Delta\sum_{i=2}^{i}\delta_{i}
\]
which can be be obtained by summing up the equations (\ref{MainEquations})
from $2$ to $i$. Then we have 
\begin{equation}
\delta_{i}=(1+Q_{i})^{-\frac{1}{2}}-1=\sum_{m=1}^{\infty}\delta_{i,m},\delta_{i,m}=a_{m}Q_{i}^{m}\label{series_main}
\end{equation}
where 
\[
Q_{i}=\Delta^{2}\sum_{i=2}^{i}F(x_{i})
\]
\begin{equation}
(-1)^{m}a_{m}=\frac{1.3...(2m-1)}{2^{m}m!}=\frac{(2m)!}{(2^{m}m!)^{2}}\sim\frac{1}{\sqrt{\pi m}},|a_{m}|\leq1\label{a_m}
\end{equation}
We see that for non-constant function $F$, this system is strongly
non-linear and the equations are intertwined.

\section{Finite differences }

Let the function $g_{i}=g(i),i\in Z,$ be given on the lattice $Z$.
Call 
\begin{equation}
\nabla g(i)=\nabla^{+}g(i)=g(i+1)-g(i),(\nabla^{-}g)(i)=g(i)-g(i-1)\label{diff_2}
\end{equation}
its right and left finite difference (discrete derivative). If the
function is defined on the part of the lattice, for example only for
$i=1,...,N$, then we consider only such differences which are defined.
For example, the difference $\nabla^{l}g(i)$ is defined iff $i+l\leq N$.

Note that

\begin{equation}
\nabla{}^{n}=(S-1)^{n}=\sum_{k=0}^{n}C_{n}^{k}(-1)^{k}S^{n-k}=\sum_{k=0}^{n}C_{n}^{k}(-1)^{n-k}S^{k}\label{diff_shift}
\end{equation}
where $S$ is the shift operator 
\[
(Sf)(i)=f(i+1)
\]
It follows in particular 
\begin{equation}
\nabla x_{i}=\Delta_{i}=x_{i+1}-x_{i},\nabla{}^{2}x_{i}=x_{i+2}-2x_{i+1}+x_{i}=\nabla\Delta_{i}=\Delta\nabla\delta_{i}=\Delta(\delta_{i+1}-\delta_{i})\label{x_i_delta_i}
\end{equation}
Thus the derivatives commute with $S$ and the following (Leibniz)
formulas hold 
\begin{equation}
\nabla(gf)(i)=f(i+1)(\nabla g)(i)+g(i)(\nabla f)(i)=(Sf)(\nabla g)+g(\nabla f)==(Sg)(\nabla f)+f(\nabla g)\label{productDiff}
\end{equation}

\begin{equation}
\nabla(f_{1}...f_{n})=(\nabla f_{1})S(f_{2}...f_{n})+f_{1}\nabla(f_{2}...f_{n})=...=\sum_{k=1}^{n}f_{1}...f_{k-1}(\nabla f_{k})S(f_{k+1}...f_{n})\label{diff_n_l}
\end{equation}
In the continuous case the differentiation of the product formula
(with arbitrary $l$ and $n$) is 
\begin{equation}
(f_{1}...f_{n})^{(l)}=\sum_{Q}\frac{l!}{\prod_{k=1}^{n}l_{k}!}f_{1}^{(l_{1})}...f_{n}^{(l_{n})}\label{diff_contin}
\end{equation}
where the sum is over all arrays $Q=(l_{1},...,l_{n})$ of non-negative
integers, satisfying the condition $l_{1}+...l_{n}=l$, and index
$(l)$ indicates the derivative of the order $l$. In the discrete
case we will need the analog of the formula (\ref{diff_contin}) 
\[
\nabla^{l}(f_{1}f_{2})=\sum_{k=0}^{l}C_{l}^{k}(\nabla^{k}f_{1})(\nabla^{l-k}S^{k}f_{2})
\]
\begin{equation}
\nabla^{l}(f_{1}...f_{n})=\sum_{Q}\frac{l!}{\prod_{k=1}^{n}l_{k}!}(\nabla^{l_{1}}f_{1})(\nabla^{l_{2}}S^{l_{1}}f_{1})(\nabla^{l_{3}}S^{l_{1}+l_{2}}f_{1})...(\nabla^{l_{n}}S^{l_{1}+...+l_{n-1}}f_{n})\label{Leibnitz_general}
\end{equation}
Both formulas are easily proved by induction, the first one is the
formula (48) in \cite{Spiegel}. However, shift operators and their
powers will not play role for us. If one denotes 
\[
\gamma_{k}(q)=\max_{i=1,...,l-q}|S^{i}\nabla^{q}f_{k}|
\]
then in two last sections we will need the inequality 
\begin{equation}
\nabla^{l}(f_{1}...f_{n})\leq\sum_{Q}\frac{l!}{\prod_{k=1}^{n}l_{k}!}\gamma_{1}(l_{1})...\gamma_{n}(l_{n})\label{Leibnitz_bound}
\end{equation}

Note also that if the function $f(i)$ does not depend on $i$, then
its differentiation gives zero. It is also easy to show that 
\begin{equation}
\nabla{}^{l}i^{n}=0,l>n\label{diff_n_bolshe_l}
\end{equation}

\paragraph{Higher differences of the power function}

The numbers 
\[
S(n,l,i)=\frac{1}{l!}\nabla^{l}i^{n}
\]
are sometimes called generalized Stirling numbers of the second kind
\cite{MedvedevIvchenko}. A particular case is the ordinary Stirling
numbers of the second kind

\[
\{\begin{array}{c}
n\\
l
\end{array}\}=S(n,l)=S(n,l,0)=\frac{1}{l!}\nabla^{l}0^{n}=\frac{1}{l!}\sum_{k=0}^{l}(-1)^{l-k}C_{l}^{k}k^{n}
\]
where the value $\nabla{}^{l}i^{n}$ for $i=0$ is denoted by $\nabla{}^{l}0^{n}$.
It is well-known that 
\begin{equation}
\{\begin{array}{c}
n\\
l
\end{array}\}=0,n<l,\{\begin{array}{c}
n\\
l
\end{array}\}=1,n=l,\label{stirling_0_1}
\end{equation}
Only some asymptotics for $S(n,l,i)$ are known: 
\begin{enumerate}
\item Riordan asymptotics for $n\to\infty,l=const$ 
\[
S(n,l)\sim\frac{l^{n}}{l!}
\]

\item for $l\to\infty,n=l+k$ with bounded $k$, for any $i$ one has \cite{MedvedevIvchenko}
\begin{equation}
S(n,l,i)\sim C_{l+k}^{k}(i+\frac{l}{2})^{k}\label{asymp_k_finite}
\end{equation}
the particular case is the result of \cite{Hsu} for $S(n,l)$; 
\item if $n\to\infty,l\to\infty$ so that $\kappa=\frac{k}{l}$ is bounded
away from zero and infinity. This is the result of Good \cite{Good}
for $i=0$, which was generalized in \cite{MedvedevIvchenko} for
$i$'s with not too fast growth. 
\end{enumerate}

\paragraph{Discretization of smooth functions}

The following example demonstrates what one can expect in case of
analytic function discretization. If some function $g(x)$ is defined
on the circle $S_{1}$, then it is equivalent to the function on all
$\text{R},$ periodic with period $1$. Example is $g(x)=\sin2\pi x$.
Put 
\begin{equation}
g_{i}=\sin\frac{2\pi i}{N},i=1,...,N\label{explicit_function}
\end{equation}
\[
g_{i,n}=\nabla^{n}g_{i},\Delta=\frac{1}{N}
\]
Then 
\[
g_{i,1}=\nabla g_{i}=g_{i+1}-g_{i}=\int_{x_{i}}^{x_{i}+\Delta}g^{(1)}(y_{1})dy_{1},
\]
\[
g_{i,2}=\nabla g_{i,1}=g_{i+1,1}-g_{i,1}=\int_{x_{i+1}}^{x_{i+1}+\Delta}g^{(1)}(y)dy-\int_{x_{i}}^{x_{i}+\Delta}g^{(1)}(y)dy=
\]
\[
=\int_{x_{i}}^{x_{i}+\Delta}(g^{(1)}(y_{1}+\Delta)-g^{(1)}(y_{1}))dy_{1}=\int_{x_{i}}^{x_{i}+\Delta}dy_{1}\int_{y_{1}}^{y_{1}+\Delta}g^{(2)}(y_{2})dy_{2},
\]
and so on 
\[
g_{i,n+1}=\nabla^{n+1}g_{i}=g_{i+1,n}-g_{i,n}=\int_{x_{i}}^{x_{i}+\Delta}dy_{1}(\int_{y_{1}}^{y_{1}+\Delta}dy_{2}....(\int_{y_{n}}^{y_{n}+\Delta}dy_{n+1}g^{(n+1)}(y_{n+1})))
\]
That is why, if $|g^{(n)}(x)|\leq C^{n+1}$ for some constant $C=C(g)>0$,
then 
\[
|g_{i,n}|\leq C^{n+1}\Delta^{n}
\]

\section{Constant external force}

\begin{theorem}Let $F(x)=F$ be constant, then as $N\to\infty$ and
$1\leq l=o(N)$ uniformly in $i<N-l$ 
\[
\nabla{}^{l+1}x_{i}\sim(-1)^{l}\sqrt{2}(\frac{F}{e}l\Delta^{2})^{l}\Delta
\]

\end{theorem}

Proof. Due to (\ref{x_i_delta_i}) it is sufficient to prove that
\[
\nabla{}^{l}\delta_{i}\sim\sqrt{2}(\frac{F}{e}l\Delta^{2})^{l}
\]
We have from (\ref{series_main}) 
\begin{equation}
\nabla{}^{l}\delta_{i+1}=\sum_{m=1}^{\infty}\nabla{}^{l}\delta_{i+1,m},\delta_{i+1,m}=a_{m}i^{m}\Delta^{2m}F^{m}\label{series_delta_l_m}
\end{equation}
By (\ref{diff_n_bolshe_l}) or (\ref{stirling_0_1}) for $l>m$ 
\[
\nabla{}^{l}\delta_{i,m}=0
\]

\begin{lemma}

For $l=m$ 
\[
\nabla{}^{l}\delta_{i+1,m}=a_{l}l!(F\Delta^{2})^{l}
\]
\end{lemma}

Note that by (\ref{diff_shift}) 
\begin{equation}
\nabla{}^{l}i^{l+p}=\sum_{k=0}^{l}C_{l}^{k}(-1)^{l-k}(i+k)^{l+p}\label{nabla_l_l+p}
\end{equation}
Then lemma follows from the chain of equalities 
\[
\nabla{}^{m}i^{m}=\sum_{k=0}^{m}C_{m}^{k}(-1)^{m-k}(i+k)^{m}=\sum_{k=0}^{m}C_{m}^{k}(-1)^{m-k}k{}^{m}=m!\{\begin{array}{c}
m\\
m
\end{array}\}=m!
\]
where the second equality has two explanations: either using (\ref{stirling_0_1})
or because any differentiation lowers by $1$ the degree of polynomial
of $i$, that is why the second expression does not depend on $i$.

To prove the theorem we will show that the sum of the series $(\ref{series_delta_l_m})$
for $m>l$ is asymptotically less than the main term with $l=m$.

Further on $m=l+p,p>0$. We consider two cases. Firstly let $p>l$,
then from (\ref{nabla_l_l+p}) it follows that 
\[
|(\nabla^{+})^{l}\delta_{i+1,l+p}|\leq|a_{l+p}|(F\Delta^{2})^{l+p}2^{l}N^{l+p}=|a_{l+p}|2^{l}F^{2l}\Delta^{2l}(F\Delta)^{p-l}
\]
Take minimal $l_{0}$, such that for any $l\geq l_{0}$ 
\[
l!>2^{l}F^{l}
\]
Then for $l\geq l_{0}$ 
\[
|a_{l+p}|2^{l}F^{2l}\Delta^{2l}(F\Delta)^{p-l}\leq|a_{l}|(F\Delta^{2})^{l}(F\Delta)^{p-l}l!\leq(F\Delta)^{p-l}|\nabla{}^{l}\delta_{i+1,l}|
\]
that means that the sum over $p>l$ of the terms with $l\geq l_{0}$
is (asymptotically) majorized by the \textquotedbl{}main\textquotedbl{}
term. If $l<l_{0}$ then the corresponding term does not exceed 
\[
C(l_{0},F)\Delta^{p-l}|\nabla{}^{l}\delta_{i+1,l}|
\]
for some constant $C(l_{0},F)>0$.

The case $1\leq p\leq l$ is more complicated. One can easily show
the known, see for example \cite{MedvedevIvchenko}, equality 
\[
\nabla{}^{l}i^{l+p}=\sum_{q=0}^{p}C_{l+p}^{q}i^{q}(\nabla^{l}0^{l+p-q})
\]

For $l\to\infty$ we will need the fact that finite differences $\nabla^{l}i^{n}$
(and Stirling numbers) have simple combinatorial interpretation \cite{MedvedevIvchenko}.
Remind that $\nabla^{l}0^{n}$ is equal to the number of ways to place
$n$ different objects in $l$ different cells so that no cell is
empty. Similarly let we have $l+i$ different cells, $l$ of them
being marked out. Then $\nabla^{l}i^{n}$ is the number of ways to
place $n$ different objects in these cells so that no of the marked
out $l$ cells is empty. It follows that $\nabla^{l}i^{n}$ are non-negative
(if $i\geq0$) and increase with $n$.

\begin{lemma} 
\[
\nabla^{l}i^{n+1}\leq(l+i+nl)\nabla^{l}i^{n}
\]

\end{lemma}

Proof. We have

\[
\nabla^{l}i^{n}\leq\nabla^{l}i^{n+1}=(l+i)\nabla^{l}i^{n}+l\nabla^{l-1}i^{n}
\]
In fact, the inequality is evident, and the equality can be explained
as follows. Fix one element (for example the last one in some fixed
enumeration). The configuration of the other elements can be of two
types: 1) such that all $l$ marked out cells were nonempty. Then
the fixed element can be placed in $l+i$ ways, 2) exactly one of
$l$ marked out cells is empty, then the fixed element should be placed
in this free cell.

Moreover it is clear that 
\[
\nabla^{l-1}i^{n}\leq n\nabla^{l}i^{n}
\]
Lemma is proved.

As $l+i\leq N$ and $nl\leq2l^{2}=o(\Delta^{2})$, we get from the
lemma 
\[
|\frac{\nabla^{l}\delta_{i+1,n+1}}{\nabla^{l}\delta_{i+1,n}}|\leq|\frac{a_{l+1}}{a_{l}}|(l+i+nl)F\Delta^{2}=o(\Delta)
\]
and that the sum of the terms with $1\leq p<l$ is asymptotically
less than $\nabla^{l}\delta_{i+1,n}$. The theorem is proved.

Remark. The asymptotics for the case when $l$ is of order $N$, is
related to the unsolved combinatorial problem of finding the maximum
of unimodal sequence of Stirling numbers, see \cite{Ajgner}, proposition
3.30. But, for example, the estimates for $l\sim\epsilon N$ 
\[
(1-C\epsilon)<\frac{|\nabla{}^{l}\delta_{i}|}{\sqrt{2}(\frac{F}{e}l\Delta^{2})^{l}}<(1+C\epsilon)
\]
for some constant $C>0$ and sufficiently small $\epsilon>0$ follow
from the proof above.

\section{Linear external force}

\begin{theorem}Let $F(x)=\alpha x,\alpha\geq1$. Then uniformly in
$2\leq l\leq N-i$ 
\[
|\nabla{}^{l}x_{i}|\leq\Delta(C\Delta)^{l}l!
\]

\end{theorem}

Proof. By (\ref{x_i_delta_i}) it is sufficient to prove that 
\[
|\nabla{}^{l-1}\delta_{i}|\leq(C\Delta)^{l}l!
\]
Similarly to the previous section 
\begin{equation}
\nabla{}^{l-1}\delta_{i}=\sum_{m=1}^{\infty}\nabla{}^{l-1}\delta_{i,m},\delta_{i,m}=a_{m}\Delta^{2m}\alpha^{m}(\sum_{k=2}^{i}x_{k})^{m}\label{Linear_0}
\end{equation}
The estimation method of each summand will depend on the pair $m,l$.

\paragraph{Small $m,l$}

For $l=2,m=1$ 
\begin{equation}
\nabla\delta_{i,1}=a_{1}\Delta^{2}\alpha\nabla\sum_{k=2}^{i}x_{k}=a_{1}\Delta^{2}\alpha x_{i+1}\label{Linear_1}
\end{equation}
For $l=3,m=1$ 
\begin{equation}
\nabla^{2}\delta_{i,1}=a_{2}\Delta^{2}\alpha\nabla^{2}\sum_{k=2}^{i}x_{k}=a_{2}\Delta^{2}\alpha\nabla x_{i+1}\sim a_{2}\alpha\Delta^{3}\label{Linear_1_3}
\end{equation}
For $l=3,m=2$ 
\begin{equation}
\nabla^{2}\delta_{i,2}=a_{2}\Delta^{4}\alpha^{2}\nabla^{2}(\sum_{k=2}^{i}x_{k})^{2}=O(\Delta^{4})\label{Linear_2_3}
\end{equation}

\paragraph{Case $2\leq l\leq m$}

We ``honestly'' differentiate once, and for the rest $l-2$ differentiations
we use the following evident bound (which holds for any $f(i)$) 
\begin{equation}
|\nabla{}^{l-2}f(i)|\leq2^{l-2}\max|f(i)|\label{Linear_1_1}
\end{equation}
Namely, denoting $\psi_{i}=\sum_{k=2}^{i}x_{k}$, we have 
\[
|(\nabla^{+})^{l-1}\delta_{i,m}|\leq\Delta^{2m}\alpha^{m}|(\nabla^{+})^{l-2}[x_{i+1}(\psi_{i+1}^{m-1}+\psi_{i+1}^{m-2}\psi_{i}+...+\psi_{i}^{m-1}]|\leq
\]

\[
\leq\Delta^{2m}\alpha^{m}2^{l-2}m(i+1)^{m-1}\leq\alpha^{m}\Delta^{m+1}2^{l-2}m,
\]
and 
\[
\sum_{m\geq l\geq2}|(\nabla^{+})^{l-1}\delta_{i,m}|\leq2^{l-2}\sum_{m\geq l\geq2}m\alpha^{m}\Delta^{m+1}\leq l(2\alpha)^{l}\Delta^{l+1}
\]
We proved even more, namely that the formulas (\ref{Linear_1}) and
(\ref{Linear_1_3}) give asymptotics for $l=2$ and $l=3$ correspondingly.

\paragraph{Case $m<l\leq N$ }

Here the bounds are essentially more complicated. Denoting
\[
\gamma_{k}=\max_{i:i+k\leq N}|\nabla{}^{k}x_{i}|
\]
 we use (for $l\geq3$) the following inductive hypothesis 
\begin{equation}
\gamma_{k}\leq\Delta(C\Delta)^{k}k!,k=2,,,,,l-1\label{inductive_linear}
\end{equation}

Consider the first summand (that is with $m=1$) in (\ref{Linear_0}),
which is equal to 
\[
a_{1}\alpha\Delta^{2}\nabla{}^{l-2}x_{i+1}
\]
and where the result follows directly from the inductive hypothesis.

For $m\geq2$, using the formulas (\ref{Leibnitz_general}), the modulus
of the expression 
\[
\nabla{}^{l-1}(\sum_{k=2}^{i}x_{k})^{m}
\]
after $l-1$ differentiations can be estimated as

\begin{equation}
|\nabla{}^{l-1}(\sum_{k=2}^{i}x_{k})^{m}|\leq\sum_{Q}C(m,l-1|q_{0},...,q_{l-1})(\max_{i}\sum_{k=2}^{i}x_{k})^{q_{0}}(\max_{i}x_{i+1})^{q_{1}}(\max_{i}|\nabla x_{i+1}|)^{q_{2}}\prod_{k\geq3}\gamma_{k-1}^{q_{k}}\label{Linear_3}
\end{equation}
where $\sum_{Q}$ is the sum over finite ordered arrays $Q=q_{0},q_{1},...,q_{l}$
of non-negative integers such that 
\[
q_{0}+\sum_{k\geq1}q_{k}=m,\sum_{k\geq1}kq_{k}=l-1
\]
Such arrays will be called admissible. Their meaning is that exactly
$q_{0}$ factors $\sum_{k=2}^{i}x_{k}$ are not differentiated at
all, $q_{1}$ factors are differentiated exactly once after what they
become equal to $x_{i+1}$, etc., $q_{l-1}$ factors are differentiated
exactly $l-1$ times. Enumerate $m$ factors in $(\sum_{k=2}^{i}x_{k})^{m}$
from $1$ to $m$. Any of the subsequent differentiations is applied
to one of these factors, giving different summands in the formula
(\ref{diff_n_bolshe_l}). Moreover, as we take maximum in $i$, one
may not take into account the shift operators in the formula (\ref{Leibnitz_general}).

For given $Q$ consider finite enumerated arrays $\alpha=\{A_{k},k=0,...,l-1\}$
of subsets of the set $\{1,...,m\}$, such that $|A_{k}|=q_{k}$ and
\[
\{1,...,m\}=\cup A_{k}
\]
The set $A_{k}$ contain those and only those elements which have
been differentiated exactly $k$ times.

Moreover,, the sequence of $l-1$ differentiations can be subdivided
onto groups $B_{k,p},p=1,...,q_{k}$, so that 
\[
\{1,...,l-1\}=\cup_{k,p}B_{k,p},|B_{k,p}|=k
\]
The meaning of the set $B_{k,p}$ is that each differentiation from
$B_{k,p}$ is applied to the $p$-th element of the set $A_{k}$.

The constants $C(m,l-1|q_{0},...,q_{l-1})$ for admissible arrays
$Q$ are equal to the number of such partitions, and thus are equal
to 
\[
(C_{m}^{q_{0}}C_{m-q_{0}}^{q_{1}}C_{m-q_{0}-q_{1}}^{q_{2}}...)(C_{l-1}^{q_{1}}(q_{1}!))(C_{l-1-q_{1}}^{2q_{2}}\frac{(2q_{2})!}{2^{q_{2}}})...(C_{l-1-q_{1}-2q_{2}...}^{kq_{k}}\frac{(kq_{k})!}{(k!)^{q_{k}}})...=
\]
\[
=\frac{m!}{\prod_{k=0}^{l-1}q_{k}!}\frac{(l-1)!}{\prod_{k=1}^{l-1}(k!)^{q_{k}}}
\]
Let us estimate from above the moduli of the summands in (\ref{Linear_0}),
using (\ref{Linear_3}). In this we use inductive hypothesis (\ref{inductive_linear})
to estimate $\gamma_{k}$, and evident bounds for $|\sum x_{i}|,|x_{i}|,|\nabla x_{i+1}|$.
As a result we get 

\[
|\nabla{}^{l-1}\delta_{i,m}|\leq\alpha^{m}\Delta^{2m}N^{q_{0}}\Delta^{\sum_{k\geq3}q_{k}}\prod_{k\geq2}((C\Delta)^{k-1}(k-1)!)^{q_{k}}\frac{m!}{\prod_{k=0}^{l-1}q_{k}!}\frac{(l-1)!}{\prod_{k=1}^{l-1}(k!)^{q_{k}}}
\]
Finally, the power of $\Delta$ will be 
\[
2m-q_{0}+m-q_{0}-q_{1}-q_{2}+\sum_{k\geq2}(k-1)q_{k}=2m-q_{0}+m-q_{0}-q_{1}-q_{2}+l-1-q_{1}-m+q_{0}+q_{1}=
\]

\[
=2m+l-1-q_{0}-q_{1}-q_{2}\geq m+l-1
\]
and the constant $C$ will have power 
\[
l-1-q_{1}-m+q_{0}+q_{1}=l-1-m+q_{0}
\]
The final estimate is 
\[
\alpha^{m}\Delta^{m+l-1}C^{l-1-m+q_{0}}\frac{m!}{\prod_{k=0}^{l-1}q_{k}!}(l-1)!\prod_{k\geq2}\frac{((k-1)!)^{q_{k}}}{(k!)^{q_{k}}}\leq
\]

\[
\leq\alpha^{m}(C\Delta)^{l-1}(l-1)!\Delta^{2}m^{2}2^{-\frac{m}{2}}C^{-m+q_{0}}\frac{1}{\prod_{k=0}^{l-1}q_{k}!}\prod_{k\geq2}\frac{1}{k{}^{q_{k}}}
\]
In fact, as for any $k\geq2$ 
\[
k!\leq k^{k}2^{-\frac{k}{2}}
\]
then 
\[
\Delta^{m}m!\leq\Delta^{m}m^{m}2^{-\frac{m}{2}}\leq\Delta^{2}m^{2}2^{-\frac{m}{2}}
\]
Now we have only to do summation over all $Q$ 
\[
\sum_{Q}C^{q_{0}}\frac{1}{\prod_{k=0}^{l-1}q_{k}!}\prod_{k\geq2}\frac{1}{k{}^{q_{k}}}\leq(\sum_{q_{0}}\frac{C^{q_{0}}}{q_{0}!})\prod_{k=1}^{l-1}\sum_{q_{k}=0}^{m}\frac{k^{-q_{k}}}{q_{k}!}\leq e^{C}e^{\sum_{i}^{l-1}k^{-1}}\leq e^{C+\ln(l-1)}=e^{C}(l-1)
\]
after what we have the estimate of any summand (\ref{Linear_0}) for
given $m$ 
\[
(C\Delta)^{l}l!\Delta^{2}m^{2}C^{-m}e^{C}2^{-\frac{m}{2}}\alpha^{m}
\]
Summation in $m$ gives 
\[
(C\Delta)^{l}l!\Delta^{2}\sum_{m=1}^{l-1}m^{2}C^{-m}e^{C}2^{-\frac{m}{2}}\alpha^{m}
\]
Note that for $C>\frac{\alpha}{\sqrt{2}}$ 
\[
\sum_{m=1}^{l-1}m^{2}C^{-m}2^{-\frac{m}{2}}e^{C}\alpha^{m}
\]
is uniformly bounded in $l$ (and in $N$). Taking into account (\ref{Linear_0}),
we see that the theorem has been proved with the constant 
\[
C=2\alpha
\]

Remark. The case $\alpha<1$ can be considered similarly - only the
constant $C$ may change. We saw also that for $l=2$ the asymptotics
depends on $i$, but for $l=3$ it does not.

\section{Power external force}

\begin{theorem}Let $F(x)=\alpha x^{n},\alpha\geq1$. Then for any
$i$ and all $2\leq l\leq N-i$ 
\[
|\nabla{}^{l}x_{i}|\leq\Delta(\Delta C){}^{l}l!,C==2\alpha n!e^{6}
\]
\end{theorem}

Proof. Similarly to above one has to estimate 
\begin{equation}
\nabla{}^{l-1}\delta_{i}=\sum_{m=1}^{\infty}a_{m}\Delta^{2m}\alpha^{m}\nabla{}^{l-1}(\sum_{k=2}^{i}x_{k}^{n})^{m}\label{Power_1}
\end{equation}

\paragraph{Case $l=2,m=1$}

\begin{equation}
\nabla\delta_{i,1}=a_{1}\Delta^{2}\alpha\nabla\sum_{k=2}^{i}x_{k}^{n}=a_{1}\Delta^{2}\alpha x_{i+1}^{n}\label{Power_2}
\end{equation}

\paragraph{Case $2\leq l\leq m$}

We ``honestly'' differentiate once (using formula (\ref{diff_n_l})),
and for the rest $l-2$ differentiations we use the estimate (\ref{Linear_1_1}).
Namely, denoting $\psi_{i}=\sum_{k=2}^{i}x_{k}^{n}$, we have 
\[
|\nabla{}^{l-1}\delta_{i,m}|\leq\Delta^{2m}\alpha^{m}|\nabla{}^{l-2}[x_{i+1}^{n}(\psi_{i+1}^{m-1}+\psi_{i+1}^{m-2}\psi_{i}+...+\psi_{i}^{m-1}]|\leq
\]
\[
\leq\Delta^{2m}\alpha^{m}2^{l-2}m(i+1)^{m-1}\leq\alpha^{m}\Delta^{m+1}2^{l-2}m,
\]
and 
\[
2^{l-2}\sum_{m\geq l\geq2}m\alpha^{m}\Delta^{m+1}\leq l(2\alpha)^{l}\Delta^{l+1}
\]
From this and from (\ref{Power_2}) the result of the theorem follows
for $l=2$.

\paragraph{First inductive procedure }

One has to estimate 
\begin{equation}
\sum_{m=1}^{l-1}\nabla{}^{l-1}\delta_{i,m}=\sum_{m=1}^{l-1}a_{m}\alpha^{m}\Delta^{2m}\nabla{}^{l-1}(\sum_{k=2}^{i}x_{k}^{n})^{m}\label{Power_3}
\end{equation}

For this we use the inductive construction very similar to the one
used for linear external force. Together with this (for $l\geq3$)
we use the inductive assumption 
\begin{equation}
\gamma_{k}=\max_{i:i+k\leq N}|\nabla{}^{k}x_{i}|\leq\Delta(C\Delta)^{k}k!,k=2,,,,,l-1\label{induction_l-1}
\end{equation}
to get the bound for $|\nabla{}^{l-1}x_{i}^{n}|$. Similarly to the
previous section we have 
\begin{equation}
|\nabla{}^{l-1}x_{i+1}^{n}|\leq\sum_{Q}C(n,l-1|q_{0},...,q_{l})(\max_{i}x_{i+1})^{q_{0}}(\max_{i}|\nabla x_{i+1}|)^{q_{1}}\prod_{k\geq2}^{l-1}\gamma_{k}^{q_{k}}\label{n_m_1-1}
\end{equation}
where 
\[
\sum_{k\geq0}q_{k}=n,\sum_{k\geq1}kq_{k}=l-1
\]
The constants $C(n,l-1|q_{0},...,q_{l})$ for the admissible arrays
$Q$ are the smae as in the previous section 
\[
\frac{n!}{\prod_{k=0}^{l-1}q_{k}!}\frac{(l-1)!}{\prod_{k=1}^{l-1}(k!)^{q_{k}}}
\]
Using the inductive assumption, the expression (\ref{n_m_1-1}) can
be estimated from above as

\[
\Delta^{\sum_{k\geq1}q_{k}}\prod_{k\geq2}^{l-1}((C\Delta)^{k}k!)^{q_{k}}\frac{n!}{\prod_{k=0}^{l-1}q_{k}!}\frac{(l-1)!}{\prod_{k=1}^{l-1}(k!)^{q_{k}}}=
\]

\[
=\Delta^{l-1+\sum_{k\geq2}q_{k}}C^{l-1-q_{1}}\frac{n!}{\prod_{k=0}^{l}q_{k}!}(l-1)!\leq(C\Delta)^{l-1}(l-1)!n!\Delta^{\sum_{k\geq2}q_{k}}C^{-q_{1}}\frac{1}{\prod_{k=0}^{l}q_{k}!}
\]
We have to do summation over all $Q$

\[
\sum_{Q}\Delta^{\sum_{k\geq2}q_{k}}C^{-q_{1}}\frac{1}{\prod_{k=0}^{l-1}q_{k}!}\leq\sum_{q_{0}}\frac{1}{q_{0}!}\sum_{q_{1}}\frac{C^{-q_{1}}}{q_{1}!}\prod_{k=2}^{l-1}\sum_{q_{k}=0}^{l-1}\frac{\Delta^{q_{k}}}{q_{k}!}\leq e^{1+c^{-1}+)l-3)\Delta}
\]
Finally we get 
\begin{equation}
|\nabla{}^{l-1}x_{i+1}^{n}|\leq(C\Delta)^{l-1}(l-1)!n!e^{2+C^{-1}}\label{induction_1-1}
\end{equation}

\paragraph{Case $1=m<l\leq N$}

Similarly, we estimate the summand with  $m=1$, using 
\[
|\nabla{}^{l-1}\delta_{i,1}|\leq a_{1}\alpha\Delta^{2}|\nabla{}^{l-1}(\sum_{k=2}^{i}x_{k}^{n})|=a_{1}\alpha\Delta^{2}|\nabla{}^{l-2}x_{i+1}^{n}|\leq
\]
\begin{equation}
\leq a_{1}\alpha\Delta^{2}\sum_{Q}C(n,l-2|q_{0},...,q_{l-2})(\max_{i}x_{i+1}\}^{q_{0}}(\max_{i}|\nabla x_{i+1}|)^{q_{1}}\prod_{k\geq2}^{l-2}(\gamma_{k})^{q_{k}}\label{n_m_1}
\end{equation}
where 
\[
\sum_{k\geq0}q_{k}=n,\sum_{k\geq1}kq_{k}=l-2
\]
The constants $C(n,l-2|q_{0},...,q_{l-2})$ for the admissible $Q$'s
are the same as before 
\[
\frac{n!}{\prod_{k=0}^{l-2}q_{k}!}\frac{(l-2)!}{\prod_{k=1}^{l-2}(k!)^{q_{k}}}
\]
Using the inductive assumption (\ref{induction_1-1}), the modulus
of the expression (\ref{n_m_1}) in the sum can be estimated as follows

\[
\alpha\Delta^{2}\Delta^{\sum_{k\geq1}q_{k}}\prod_{k\geq2}^{l-2}((C\Delta)^{k}k!)^{q_{k}}\frac{n!}{\prod_{k=0}^{l-2}q_{k}!}\frac{(l-2)!}{\prod_{k=1}^{l-2}(k!)^{q_{k}}}=
\]

\[
=\alpha\Delta^{2+l-2+\sum_{k\geq2}q_{k}}C^{l-2-q_{1}}\frac{n!}{\prod_{k=0}^{l-2}q_{k}!}(l-2)!\leq\alpha(C\Delta)^{l}l!n!\frac{1}{l(l-1)}\Delta^{\sum_{k\geq2}q_{k}}C^{-2-q_{1}}\frac{1}{\prod_{k=0}^{l-2}q_{k}!}
\]
Again, we have to do summation over all $Q$

\[
\sum_{Q}\Delta^{\sum_{k\geq2}q_{k}}C^{-q_{1}}\frac{1}{\prod_{k=0}^{l-2}q_{k}!}\leq\sum_{q_{0}}\frac{1}{q_{0}!}\sum_{q_{1}}\frac{\Delta^{q_{1}}C^{-q_{1}}}{q_{1}!}\prod_{k=2}^{l-2}\sum_{q_{k}=0}^{l-2}\frac{\Delta^{q_{k}}}{q_{k}!}\leq e^{1+C^{-1}+(l-4)\Delta}
\]
As the result we get 
\[
\alpha(C\Delta)^{l}l!n!\frac{1}{l(l-1)}\frac{e^{2+C^{-1}}}{C^{2}}
\]

\paragraph{Second inductive procedure}

Now, using the estimate (\ref{induction_1-1}), we consider the case
$2\leq m<l\leq N$, that is estimate for $m\geq2$ the expression
\[
a_{m}\alpha^{m}\Delta^{2m}\nabla{}^{l-1}(\sum_{k=2}^{i}x_{k}^{n})^{m}
\]
Denote now 
\[
\beta_{k}=\max_{i}|\nabla{}^{k}x_{i+1}^{n}|
\]
It can be written, after $l-1$ differentiations, as 
\begin{equation}
a_{m}\alpha^{m}\Delta^{2m}\sum_{Q}C(m,l-1|q_{0},...,q_{l-1})(\max_{i}\sum_{k=2}^{i}x_{k}^{n})^{q_{0}}(\max_{i}x_{i+1}^{n})^{q_{1}}\prod_{k\geq2}^{l-1}\beta_{k-1}^{q_{k}}\label{Power_4}
\end{equation}
Finally we have 
\[
|\nabla{}^{l-1}\delta_{i,m}|=|a_{m}\Delta^{2m}\alpha^{m}\nabla{}^{l-1}(\sum_{k=2}^{i}x_{k}^{n})^{m}|\leq
\]
\[
\leq|\alpha^{m}\Delta^{2m}N^{q_{0}}\Delta^{\sum_{k\geq3}q_{k}}\prod_{k\geq2}((C\Delta)^{(k-1)}(k-1)!n!e^{2+C^{-1}})^{q_{k}}\frac{m!}{\prod_{k=0}^{l-1}q_{k}!}\frac{(l-1)!}{\prod_{k=1}^{l-1}(k!)^{q_{k}}}|
\]
Now it will be more convenient to consider the factors separately.

The power of $\Delta$ is 
\[
2m-q_{0}+\sum_{k\geq3}q_{k}+l-q_{1}-m+q_{0}+q_{1}=m+l+\sum_{k\geq3}q_{k}
\]
the power of $C$ is 
\[
l-q_{1}-m+q_{0}+q_{1}=l-m+q_{0}
\]
The remaining constants are 
\[
(n!e^{2+C^{-1}})^{m-q_{0}-q_{1}}
\]
Besides the factor $(C\Delta)^{l}l!$ we have the factor 
\[
m!l^{-1}\Delta^{m}\leq l^{-1}
\]
The summation gives 
\[
\sum_{Q}[\sum_{q_{0}=0}^{l-1}\frac{1}{q_{0}!}][\sum_{q_{1}=0}^{l-1}\frac{1}{q_{1}!}]\prod_{k=2}^{l-1}[\sum_{q_{k}=0}^{l-1}\frac{1}{q_{k}!}\frac{1}{k{}^{q_{k}}}\Delta^{q_{k}}]\leq e^{3}
\]
Finally we get the constants 
\[
C^{-m+q_{0}}(\alpha n!e^{2+C^{-1}})^{m-q_{0}-q_{1}}e^{3}
\]
which proves the theorem with 
\[
C=2\alpha n!e^{6}
\]

\end{document}